\documentclass[aps,twocolumn,floatfix,superscriptaddress, nobibnotes,longbibliography]{revtex4-2} 
\usepackage{amsmath,amssymb,physics}
\usepackage[hypertexnames=false]{hyperref}
\usepackage{datetime}
\usepackage{enumitem}
\usepackage{graphicx}
\usepackage{braket}
\usepackage{esdiff}
\usepackage{MnSymbol}
\usepackage[normalem]{ulem}

\usepackage{siunitx}
\usepackage[dvipsnames]{xcolor}
\usepackage{xfrac} 
\usepackage{tikz}

\def\kup{\left\vert\uparrow\right\rangle}
\def\kdn{\left\vert\downarrow\right\rangle}
\def\nch{n_\mathrm{m}}
\renewcommand{\vec}{\mathbf}

\newcommand{\new}[1]{{#1}}
\newcommand{\newnew}[1]{{#1}}
\hypersetup{
    colorlinks,
    linkcolor={red!50!black},
    citecolor={blue!50!black},
    urlcolor={blue!80!black}
}

\footnotetext{These authors contributed equally to this work}

\begin{document}
\makeatletter
\newcommand{\fmarki}{\ensuremath{\dagger}}
\def\@fnsymbol#1{{\ifcase#1\or \fmarki \else\@ctrerr\fi}}
\makeatother

\title{Superfluid signatures in a dissipative quantum point contact}

\author{Meng-Zi Huang\textsuperscript{*}}
\email{mhuang@phys.ethz.ch}

\author{Jeffrey Mohan\textsuperscript{*}}

\affiliation{Institute for Quantum Electronics, ETH Z\"urich, 8093 Z\"urich, Switzerland}
\author{Anne-Maria Visuri}
\affiliation{Physikalisches Institut, University of Bonn, Nussallee 12, 53115 Bonn, Germany}
\author{Philipp Fabritius}
\author{Mohsen Talebi}
\author{Simon Wili}
\affiliation{Institute for Quantum Electronics, ETH Z\"urich, 8093 Z\"urich, Switzerland}
\author{Shun Uchino}
\affiliation{Advanced Science Research Center, Japan Atomic Energy Agency, Tokai 319-1195, Japan}
\author{Thierry Giamarchi}
\affiliation{Department of Quantum Matter Physics, University of Geneva, 24 quai Ernest-Ansermet, 1211 Geneva, Switzerland}
\author{Tilman Esslinger}
\affiliation{Institute for Quantum Electronics, ETH Z\"urich, 8093 Z\"urich, Switzerland}

\date{\today}

\begin{abstract}
We measure superfluid transport of strongly interacting fermionic lithium atoms through a quantum point contact with local, spin-dependent particle loss. We observe that the characteristic non-Ohmic superfluid transport enabled by high-order multiple Andreev reflections transitions into an excess Ohmic current as the dissipation strength exceeds the superfluid gap. We develop a model with mean-field reservoirs connected via tunneling to a dissipative site. Our calculations in the Keldysh formalism reproduce the observed nonequilibrium particle current, yet do not fully explain the observed loss rate or spin current.
\end{abstract}

\maketitle

The interplay between coherent Hamiltonian dynamics and incoherent, dissipative dynamics emerging from coupling to the environment leads to rich phenomena in open quantum systems \cite{breuer2002,daley_quantum_2014, ashida_non-hermitian_2020}, including the quantum Zeno effect \cite{barontini_controlling_2013, zhu_suppressing_2014,tomita_observation_2017, froeml_diehl_2019, dolgirev_non-gaussian_2020, will_controlling_2022}, emergent dynamics \cite{miri_exceptional_2019, letscher_bistability_2017, dogra_dissipation-induced_2019, bouganne_anomalous_2020, dreon_self-oscillating_2022, wu_dynamical_2022}, and dissipative phase transitions \cite{kessler_dissipative_2012, fink_signatures_2018, helmrich_signatures_2020, ferri_emerging_2021, yamamotoDissipativeFermionicSuperfluid2021, benary_experimental_2022}. Moreover, an important question is how many-body coherence competes with dissipation by dephasing or particle loss. Directed transport between two reservoirs offers an advantageous setup for studying this competition since dissipation can be applied locally without perturbing the many-body states in the reservoirs \cite{amico_roadmap_2021}. So far, studies on dissipation in solid-state systems have focused on dephasing \cite{schon_quantum_1990, penttila_superconductor-insulator_1999, murani_absence_2020}. More recently, quantum gases have become versatile platforms to study interacting many-body physics and to engineer novel forms of dissipation \cite{daley_quantum_2014, barontini_controlling_2013}, though previous transport experiments on dissipation have focused on weakly interacting systems \cite{labouvie_bistability_2016, corman_quantized_2019, gou_tunable_2020}.

\begin{figure}[tb]
    \includegraphics{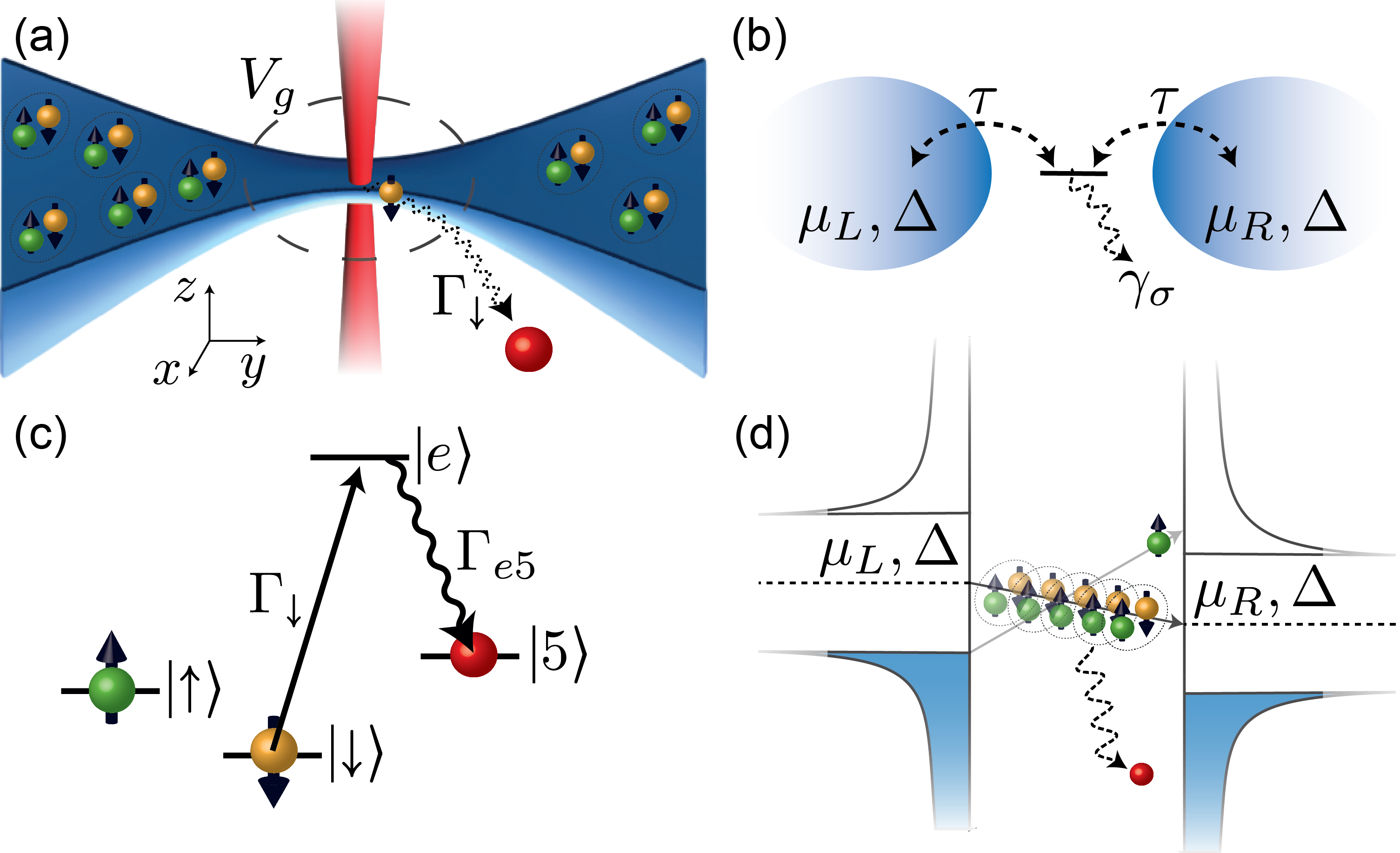}
    \caption{(a) Two-terminal transport setup of a strongly interacting Fermi gas with a dissipation beam resonant with $\kdn$ at the center of a 1D channel. A gate beam (dashed circle) exerts an attractive potential $V_g$ that determines the superfluid gap $\Delta$ and the number of transport modes $\nch$. (b)~Theoretical model where the channel is modeled by a single, lossy site tunnel coupled to BCS reservoirs. (c) Dissipation scheme showing the relevant atomic energy levels. $\kdn$ is optically excited by the dissipation beam to $\ket{e}$ which decays predominantly to an auxiliary ground state $\ket{5}$ that quickly leaves the system. (d) Illustration of MAR in a superconducting QPC: transporting a quasiparticle requires energy to overcome the gap $2\Delta$, which is enabled by cotunneling of many pairs, each providing a small energy $2(\mu_L-\mu_R)$. The quasiparticle and constituents of the pairs experience dissipation, inhibiting this process.}
    \label{fig:fig1}
\end{figure}

Engineered dissipation in strongly correlated fermionic systems, while only starting to be explored theoretically \cite{damanet_controlling_2019, yamamotoDissipativeFermionicSuperfluid2021}, opens interesting themes such as its competition with superfluidity where pairing and many-body coherence are key. While the Josephson effect is an archetype of superfluid transport, irreversible currents between superfluids are highly nontrivial but less studied in cold-atom systems.
A prime example is the excess current between two superconductors~\cite{bretheau_superconducting_2012} or superfluids~\cite{Husmann_quantum_point_contact2015} through a high-transmission quantum point contact (QPC) under a chemical potential bias where the Josephson current is suppressed. 
Because of the superfluid gap $\Delta$, direct quasiparticle transport is suppressed when the chemical potential difference $\Delta\mu$ between the reservoirs is smaller than $2\Delta$ [illustrated in Fig.~\ref{fig:fig1}(d)]. Instead, this energy barrier can be overcome by cotunneling of many Cooper pairs $n_\mathrm{pair}\geq\Delta/\Delta\mu$, each providing an energy $2\Delta\mu$ \cite{cron_multiple-charge-quanta_2001, cuevas_full_2003}. This process is known as multiple Andreev reflections (MAR)~\cite{blonder_transition_1982, averin_josephson_1995, cuevas_hamiltonian_1996, bolech_keldysh_2005}. The robustness of MAR to dissipation is an interesting open question, especially for pair-breaking particle loss acting on only one spin state, since the very existence of MAR relies on many-body coherence between the spins.

In this work, we address this question by experimentally and theoretically studying the influence of spin-dependent particle loss on superfluid transport. We use a strongly correlated Fermi gas---a superfluid with many-body pairing---in a transport setup with two reservoirs connected by a QPC  
and apply controllable local particle loss at the contact. We find that, surprisingly, the superfluid behavior survives for dissipation strengths larger than $\Delta$---the energy scale responsible for the observed current.
This result is reproduced by a minimal model that includes both superconductivity and dissipation written in the Keldysh formalism.

\emph{Experiment.}---We prepare a degenerate Fermi gas of $^6$Li in a harmonic trap in a balanced mixture of the first- and third-lowest hyperfine ground states, labeled $\kdn$ and $\kup$. The atomic cloud has typical total atom numbers $N=N_\downarrow+N_\uparrow=195(14)\times10^3$, temperatures $T=\SI{100(2)}{nK}$, and Fermi temperatures 
$T_F=h\bar{\nu}_\mathrm{trap}(3N)^{1/3}/k_B=391(10)$\,nK, where $h$ is the Planck constant, $k_B$ the Boltzmann constant, and $\bar{\nu}_\mathrm{trap}=98(2)$\,Hz the geometrical mean of the harmonic trap frequencies.
Using a pair of repulsive, $\text{TEM}_{01}$-like beams intersecting at the center of the cloud, we optically define two half-harmonic reservoirs connected by a quasi-1D channel with transverse confinement frequencies $\nu_x=\SI{10(2)}{kHz}$ and $\nu_z=\SI{9.9(2)}{kHz}$, realizing a QPC illustrated in Fig.~\ref{fig:fig1}(a). We apply a magnetic field of 689.7\,G to address the spins' Feshbach resonance, giving rise to a fermionic superfluid in the densest parts of the cloud at the contacts to the 1D channel. 
An attractive Gaussian beam propagating along $z$ acts as a ``gate'' potential $V_g$ which increases the local chemical potential at the contacts. 
It enhances the local degeneracy $T/T_F\sim 0.03$ \cite{supplementary} well into the superfluid phase and determines both the superfluid gap $\Delta$ and the number $n_\mathrm{m}$ of occupied transverse transport modes in the contact.
Both $\Delta$ and $\nch$ can be computed from the known potential energy landscape and equation of state \cite{supplementary}
\nocite{breuer2002}
\nocite{Husmann_quantum_point_contact2015}
\nocite{setiawan_analytic_2022}
\nocite{martin2011josephson}
\nocite{visuri2022}
\nocite{kamenev2011, sieberer2016}
\nocite{cuevas_hamiltonian_1996}
\nocite{jinGenericTransportFormula2020, uchino_comparative_2022}
\nocite{lu_spin-polarized_2020}
\nocite{martin-rodero_microscopic_1996}
\nocite{tinkham_introduction_1996}
\nocite{likharev_superconducting_1979}
\nocite{del_pace_tunneling_2021}
\nocite{meier_josephson_2001}
\nocite{valtolina_roati2015}
\nocite{krinnerObservationQuantizedConductance2014}
\nocite{yao_controlled_2018}
\nocite{uchino_bosonic_2020, uchino_role_2020}
\nocite{husmannBreakdownWiedemannFranz2018}
\nocite{lebratQuantizedConductanceSpinSelective2019}
\nocite{corman_quantized_2019}
\nocite{houbiers_elastic_1998}
\nocite{zwergerBCSBECCrossoverUnitary2012}
\nocite{thomas_virial_2005}
\nocite{shinPhaseDiagramTwocomponent2008, olsen_phase_diagram_2015}
\nocite{rammelmuller_finite-temperature_2018}
\nocite{kuRevealingSuperfluidLambda2012}
\nocite{hou_first_2013}
\nocite{zurn_precise_2013}
\nocite{long_spin_susceptibility_2021,rammelmuller_pairing_2021}
\nocite{boettcher_phase_2015}
\nocite{scheerSignatureChemicalValence1998}
\nocite{ihnSemiconductorNanostructuresQuantum2010}
\nocite{kanasz-nagy_anomalous_2016}
\nocite{scheer_conduction_1997}
\nocite{schirotzek_2008}
and are approximately $\Delta \approx k_B \times \SI{1.4}{\micro K} \approx \hbar\times \SI{184}{ms^{-1}}$ and $\nch \approx 3$ in this work.

We engineer spin-dependent particle loss with a tightly focused beam at the center of the 1D channel [Fig.~\ref{fig:fig1}(a)] that optically pumps $\kdn$ to an auxiliary ground state $\ket{5}$ [Fig.~\ref{fig:fig1}(c)] which interacts weakly with the two spin states and is lost due to photon recoil \cite{supplementary}. This leads to a controllable particle dissipation rate of $\kdn$ atoms given by the peak photon scattering rate $\Gamma_\downarrow$ with no observable heating in the reservoirs. While this loss beam is far off-resonant for $\kup$, causing no dissipation in the absence of interatomic interactions, we observe loss of $\kup$ in the strongly-interacting regime studied here \cite{supplementary}.
Even at the strongest dissipation, the system lifetime is over a second, much longer than the timescale of a detectable transport of $10^3$ atoms via MAR $\sim10^3h/\Delta \sim \SI{30}{ms}$.

\begin{figure}[tb]
    \includegraphics{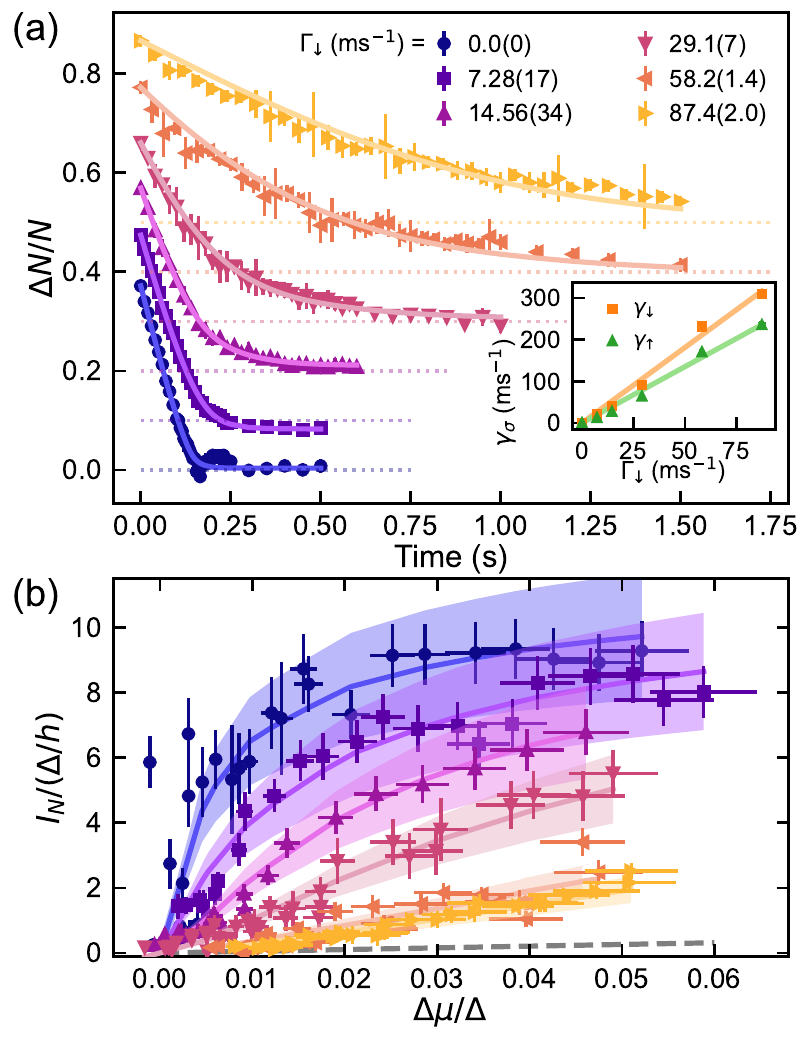}
    \caption{\newnew{(a) Time evolution of particle imbalance for various dissipation strengths $\Gamma_\downarrow$. Solid curves are fits of the theoretical model and the data sets are vertically offset by 0.1 for clarity. Error bars represent $1\sigma$ statistical uncertainty of 4 to 5 repetitions. The fitted dissipation $\gamma_\sigma$ ($\gamma_\uparrow\equiv 0.72\gamma_\downarrow$) for each curve is plotted versus $\Gamma_\downarrow$ in the inset, showing a linear relation (solid lines are linear fits). (b) Numerically extracted current-bias relation for the same data sets in units of the gap $\Delta\approx \hbar\times \SI{184}{ms^{-1}}$. The dashed line indicates the maximum normal conductance $2\nch/h$. Error bars include uncertainties in the conversion to reduced units. The uncertainties in the calculated current, represented by the shaded bands, originate mostly from the number of transmission modes $\nch$.}}
    \label{fig:fig2}
\end{figure}

We induce particle transport from the left to the right reservoir by preparing an atom number imbalance $\Delta N = N_L-N_R$ that generates a chemical potential bias $\Delta\mu = \Delta\mu(\Delta N,N,T)$ given by the system's equation of state \cite{supplementary}. $\Delta\mu$ drives a current $I_N=-\dot{\Delta N}/2$ from left to right that causes $\Delta N$ to decay over time $t$.
The dynamics of $\Delta N(t)/N(t)$ for various $\Gamma_\downarrow$ are plotted in Fig.~\ref{fig:fig2}(a). From each trace, we numerically extract the current-bias relation $I_N(\Delta\mu)$, shown in Fig.~\ref{fig:fig2}(b) in units of the superfluid gap $\Delta$ \cite{supplementary}.

For weakly interacting spins, the decay of $\Delta N(t)$ at arbitrary $\Gamma_\downarrow$ is exponential since a QPC coupling two Fermi liquids has a linear (Ohmic) current-bias relation $I_N=G\Delta\mu$. The conductance $G$ is quantized in units of $2/h$ (2 comes from spin) but can be renormalized to a smaller value by dissipation \cite{lebratQuantizedConductanceSpinSelective2019, corman_quantized_2019}. In contrast, we observe that the decay of $\Delta N$ at $\Gamma_\downarrow=0$ deviates strongly from an exponential and the corresponding current-bias relation $I_N(\Delta\mu)$ is highly nonlinear, consistent with previous observations in the strongly interacting regime \cite{Husmann_quantum_point_contact2015}. In fact, this nonlinearity is a signature of superfluidity. Specifically, for ballistic superconducting QPCs \cite{cuevas_hamiltonian_1996}, the current is approximately $I_N \approx G\Delta\mu + I_N^\mathrm{exc}$ where the normal current $G\Delta\mu \approx 2\nch\Delta\mu/h$ is carried by quasiparticles. The excess current $I_N^\mathrm{exc} \approx (16/3)\nch\Delta/h$ carried by the Cooper pairs is given by the superfluid gap $\Delta$---the natural energy scale in this system. The dominance of the excess current over the normal current can be seen in $\Delta N(t)/N$ with $\Gamma_\downarrow=0$: It decays almost linearly in time, indicating that the current is nearly independent of $\Delta\mu$.
We do not expect Josephson oscillations (as in Ref.~\cite{valtolina_roati2015,del_pace_tunneling_2021}) here since the irreversible current (MAR) damps the reversible Josephson current below our experimental resolution \cite{meier_josephson_2001,yao_controlled_2018}.

We now turn to the question of the influence of dissipation on the superfluid transport. At low $\Gamma_\downarrow$, superfluidity is still evident from the linear initial decay of $\Delta N$ which gives way to exponential behavior at low bias where the MAR responsible for transport become higher order.
The effect of dissipation is clearer in the extracted current-bias relations shown in Fig.~\ref{fig:fig2}(b). As dissipation strength increases, the initial current (largest $\Delta\mu$) reduces as does the concavity of the curve. The current-bias relation eventually approaches a straight line at high dissipation, yet the slope is still significantly larger than the normal conductance $2\nch/h$ \newnew{[dashed line in Fig.~\ref{fig:fig2}(b)]}. 
We argue below, \newnew{after fitting our theoretical model to the data,} that this excess conductance, albeit linear, is a signature of the MAR process. An increased temperature leading to a suppressed gap can result in similar anomalous conductance \cite{Husmann_quantum_point_contact2015, krinner_mapping_2016, kanasz-nagy_anomalous_2016, uchino_anomalous_2017, liu_anomalous_2017}, though we can exclude this effect here from the absence of temperature increase in the reservoirs.
Another possible contribution to the linear conductance in a unitary Fermi gas arises from pair tunneling coupled to collective modes in the superfluid \cite{meier_josephson_2001,uchino_role_2020}. However, it is expected to be on the order of $2/h$ per mode, thus negligible compared to the MAR contribution even at strong dissipation. This is also consistent with the small linear conductance (slope at large bias $\sim 2n_\mathrm{m}/h$) in the observed current-bias relation at $\Gamma_\downarrow=0$.

\emph{Theoretical model and fit.}---We have developed a mean-field model based on previous work \cite{Husmann_quantum_point_contact2015} by adding an atom loss process within the Lindblad master equation framework \cite{supplementary}. The two reservoirs are treated as BCS superfluids, and transport through the channel is modeled by single-particle tunneling with amplitude $\tau$ from either reservoir onto a dissipative site between them [Fig.~\ref{fig:fig1}(b)]. The dissipation is modeled with a Lindblad operator $\hat{L}_\sigma = \sqrt{\gamma_\sigma} \hat{d}_\sigma$ proportional to the site's fermionic annihilation operator $\hat{d}_\sigma$ with a rate $\gamma_\sigma$ for each spin $\sigma=\downarrow,\uparrow$, i.e.~as a pure particle loss process that is uncorrelated between the two spins. 
$\gamma_\downarrow$ is directly related to $\Gamma_\downarrow$ as our results below show, while $\gamma_\uparrow$ is included phenomenologically to match the experimental observation that $\kup$ is also lost due to the strong interaction.

To compute nonequilibrium observables such as $I_N$, we use the Keldysh formalism extended to dissipative systems~\cite{kamenev2011, sieberer2016, jinGenericTransportFormula2020, yamamotoDissipativeFermionicSuperfluid2021, visuri2022, uchino_comparative_2022, visuri_nonlinear2022}. The time integral of the theoretical current $I_N^K(\Delta\mu,\tau,\gamma_\downarrow,\gamma_\uparrow)$ along with the equation of state $\Delta\mu(\Delta N,N,T)$ yields the model's prediction for the time evolution of the particle imbalance $\Delta N^K(t,\tau,\gamma_\downarrow,\gamma_\uparrow)$. Here, $N(t)$ is obtained from an exponential fit to the data \cite{supplementary}. For simplicity, we model a single transport mode and obtain the net current by multiplying by the number of modes $\dot{\Delta N^K} = -2 \nch I_N^K$. Although the formalism can treat nonzero temperatures, we simplify the calculation by using zero temperature since $k_B T/\Delta < 0.08$.

Because of the spin-dependent loss, a spin imbalance builds up, leading to a magnetization imbalance $\Delta M = \Delta N_\downarrow - \Delta N_\uparrow \neq 0$ that can drive additional particle current. Nevertheless, $\Delta M$ remains small ($\Delta M/N < 0.07$) for all our data and its effect on $I_N$ is negligible \cite{supplementary}. Moreover, the model predicts a spin current 1 order of magnitude below the observed value, and therefore does not fully describe the spin-dependent transport. In this work, we focus on spin-averaged particle transport.

We perform a least-squares fit of $\Delta N^K(t, \tau, \gamma_\downarrow, \gamma_\uparrow)$ to the measured $\Delta N(t)$ to extract the model parameters $\tau$, $\gamma_\downarrow$, and $\gamma_\uparrow$. 
In our case of low bias and ballistic channel, $I_N^K$ is sensitive only to the sum $\gamma = \gamma_\uparrow + \gamma_\downarrow$. Therefore, to avoid overfitting, we fix the ratio $\gamma_\uparrow/\gamma_\downarrow=r$ to the average measured ratio of the atom loss rates of the two spin states $r = \dot{N}_\uparrow/\dot{N}_\downarrow=0.72(4)$ \cite{supplementary}. As the gap $\Delta$ is the natural unit of current and bias, the estimation of $\Delta$ in the experimental system is crucial for quantitative comparison to the model. However, the spatially varying gap in the potential energy landscape of the experiment (crossover from the 3D reservoirs to the 1D channel) is not explicitly modeled. Motivated by this and an experimental uncertainty of about 6\% in the potential energy of the gate beam, we fit a multiplicative correction factor $\eta_g$ on the gate potential $V_g$, which strongly affects both our estimate of $\Delta$ in the most degenerate point in the system and $\nch$ \cite{supplementary}. Since the system is identical in each dataset except for the dissipation strength, we first fit the $\Gamma_\downarrow=0$ data with $\gamma_\downarrow = \gamma_\uparrow = 0$ to find $\tau$ and $\eta_g$, which determine the concavity of $I_N^K(\Delta\mu)$ and the timescale of $\Delta N^K(t)$, respectively. We then fit the single parameter $\gamma_\downarrow$ with fixed $r$, $\tau$, and $\eta_g$ for subsequent sets. The systematic error in $\eta_g$ is the dominant source of uncertainty in our theoretical calculations.

The fits for each dataset are plotted as solid lines in Figs.~\ref{fig:fig2}(a),(b). With this fitting procedure, we find $\eta_g=0.98(6)$, $\Delta/k_B=\SI{1.4(1)}{\micro K}$, and $\nch=2.6(5)$. 
The fitted $\tau$ determines the energy linewidth of the dissipative site $\Gamma_d\propto\tau^2$ \cite{supplementary}, which in units of the gap is $\Gamma_d=5.9(4)\Delta$.
Independent measurements with different values of $\nu_x$ and $V_g$ produce similar results for $\eta_g$ and $\Gamma_d$ within 20\%. The large value of $\Gamma_d$ reflects the near-perfect transmission of the ballistic channel as the limit $\Gamma_d\rightarrow\infty$ is equivalent to perfect transparency $\alpha\rightarrow1$ in a QPC model with direct tunneling between the reservoirs \cite{supplementary, Husmann_quantum_point_contact2015, martin2011josephson}. In principle, the energy $\epsilon_d$ of the dissipative site is another free parameter. Because of the large linewidth, however, the model is insensitive to changes of $\epsilon_d$ within the physically meaningful range $\abs{\epsilon_d} < \Delta$. There are thus no resonant effects as in a weakly coupled quantum dot, and we fix $\epsilon_d = 0$. For the same reason, we do not consider any on-site interaction.

The fitted $\gamma_\downarrow$ and $\gamma_\uparrow$ versus $\Gamma_\downarrow$, plotted in the inset of Fig.~\ref{fig:fig2}(a), show that the effective dissipation strength is approximately proportional to the photon scattering rate as expected. 
The fitted slope $\gamma_\downarrow = 3.6\Gamma_\downarrow$ is of order 1, corroborating the use of our single-site model where only one $\kdn$ fermion fits into the dissipation beam at a time. In accordance, the dissipation beam's waist $w_y=\SI{1.31(2)}{\micro m}$ is comparable to the Fermi wavelength in the channel $\lambda_F \approx \SI{2}{\micro m}$.

\begin{figure}
    \includegraphics{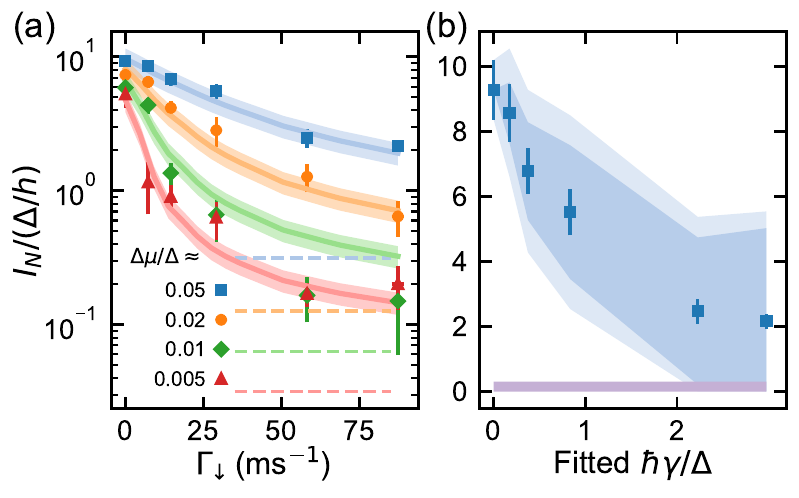}
    \caption{
    (a) Measured current vs.~dissipation strength at different biases. The theoretical calculations are shown as solid curves with uncertainties in lighter colors. Dashed lines represent upper bounds of normal-state currents $2\nch\Delta\mu/h$.
    (b) Same data at $\Delta\mu/\Delta\approx 0.05$ with bounds on the conserved current $I^\mathrm{cons} \in [I_N-\dot{N}/2, I_N+\dot{N}/2]$ indicated by the shaded area (lighter color represents the uncertainty). The horizontal axis is the fitted $\gamma=\gamma_\downarrow+\gamma_\uparrow$ in units of the gap to illustrate the effective strength of the dissipation. The conserved current is above the normal current (bounds represented by the horizontal bar), showing superfluid character even for the strongest dissipation.
    }
    \label{fig:fig3}
\end{figure}

\emph{Robustness of superfluid transport to dissipation.}---Supported by the overall fit to the data, our calculations provide insight into the observed flattening of the nonlinear current-bias relation: a plausible scenario is that dissipation suppresses higher-order MAR processes while allowing lower-order MAR ($n_\mathrm{pair}<\Delta/\Delta\mu$) to contribute \cite{supplementary}. Despite the nonlinearity disappearing as dissipation increases, the current still originates from MAR. A benchmark for this superfluid signature is the excess current above the possible normal current $2n_\mathrm{m}\Delta\mu/h$. To see this quantitatively,
we replot the measured and calculated current in Fig.~\ref{fig:fig3}(a) versus $\Gamma_\downarrow$ at a few bias values including the initial bias $\Delta\mu/\Delta\approx 0.05$ ($n_\mathrm{pair}\approx20$) down to $\Delta\mu/\Delta\approx 0.005$ ($n_\mathrm{pair}\approx200$). The fitted model is shown in solid curves using the fitted linear relationship between $\gamma_\downarrow$ and $\Gamma_\downarrow$ in the inset of Fig.~\ref{fig:fig2}(a). 
The observed current well exceeds the upper bound of normal current (dashed lines in corresponding colors) for all biases, indicating the persistence of the MAR-enabled current.
Moreover, the excess current decays smoothly with dissipation and gives no indication of a dissipative phase transition but rather shows a dissipative superfluid-to-normal crossover (cf.~Ref.\cite{tomita_observation_2017, murani_absence_2020}).

Since the current $I_N=-(\dot{N}_L-\dot{N}_R)/2$ could in principle arise purely from an asymmetric particle loss, we verify that the conserved current $I^\mathrm{cons}$---the atoms transported through the QPC without being lost---exceeds the upper bound for the normal current at dissipation strengths above the gap. This is shown in Fig.~\ref{fig:fig3}(b) where we plot the data for the largest bias together with bounds for $I^\mathrm{cons}$. These bounds are obtained by writing $\dot{N}_{L/R} = \mp I^\mathrm{cons} - I_{L/R}^\mathrm{loss}$, where $I_{L/R}^\mathrm{loss}$ is the dissipation-induced current into the vacuum \cite{uchino_comparative_2022}, and thus $I_N= I^\mathrm{cons} + (I_L^\mathrm{loss} - I_R^\mathrm{loss})/2$. Assuming the worst-case scenario---maximally asymmetric loss---leads to $I^\mathrm{cons} \in [I_N-\dot{N}/2, I_N+\dot{N}/2]$. 
The excess conserved current shows that the system preserves its superfluid signature up to $\hbar\gamma\gtrsim\Delta$.

\begin{figure}
    \includegraphics{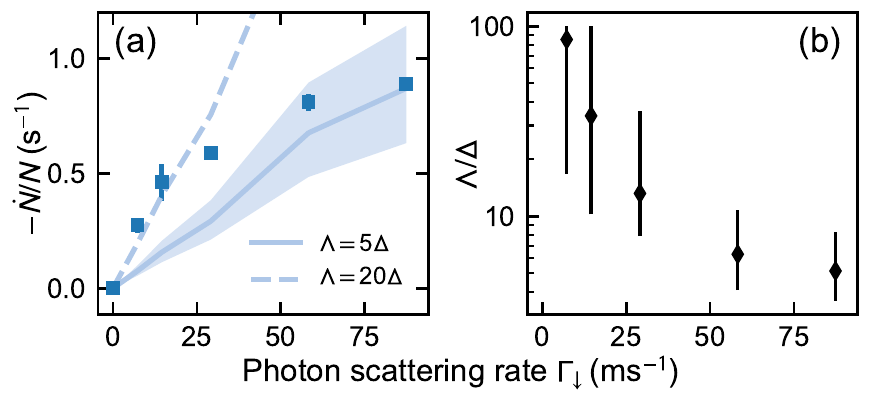}
    \caption{(a) Normalized atom loss rate vs.~dissipation strength at the initial bias, which shows saturation at high dissipation. The curves are calculations with different energy cutoff $\Lambda$. 
    Uncertainties due to $\Delta$ and $\nch$ are represented by the shaded area (only shown for the solid curve). (b) The energy cutoff needed to reproduce each data point with nonzero dissipation, indicating the underlying energy dependence neglected in our model. The error bars account for uncertainties in $\Delta$, $n_\mathrm{m}$, and $\gamma$.
    }
    \label{fig:fig4}
\end{figure}

\emph{Atom loss rate.}---Finally, we compare the theoretical model to the experiment in terms of the total atom loss rate $\dot{N}=\dot{N}_L+\dot{N}_R$. In the Keldysh formalism, computing steady-state observables involves an integration over energy. Because our model assumes reservoirs with linear dispersion,
we introduce a high energy cutoff $\Lambda$.
The current only has contributions from energies constrained by $\Delta\mu$ and $\Delta$, and is therefore independent of the cutoff when $\Lambda > \Delta, \Delta\mu$. The atom loss rate on the other hand depends on the cutoff and converges to its maximum value when $\Lambda \gg \Gamma_d$.
We find that a single $\Lambda$ cannot reproduce the observed atom loss rate at different dissipation [shown in Fig.~4(a) for $\Delta\mu/\Delta\approx0.05$]: The atom loss rate saturates as dissipation increases, while the calculated loss rate increases almost linearly with dissipation in the measured range. 
In particular, $\Lambda \gtrsim 20\Delta$ is needed to reproduce the data at weak dissipation, close to the converged value, while $\Lambda \approx 5\Delta$ reproduces the data at strong dissipation. This discrepancy is shown in Fig~\ref{fig:fig4}(b) which plots the $\Lambda$ needed to reproduce $\dot{N}$ at each measured $\Gamma_\downarrow$. Physically, this means that additional mechanisms not included in the model suppress the occupation of the dissipative site and hence the loss rate. The calculations do predict a saturation of $\dot{N}$ at an order of magnitude higher dissipation strength followed by a slow decrease of $\dot{N}$ versus $\Gamma_\downarrow$---a signature of the quantum Zeno effect \cite{barontini_controlling_2013, zhu_suppressing_2014, tomita_observation_2017}. 
We do not observe this nonmonotonicity, and expect it to be washed out due to the lack of energy resolution in $\dot{N}$ \cite{froeml_ultracold_2020} and the finite width of the dissipative beam.

\emph{Conclusion.}---We have shown a remarkable robustness of the MAR processes responsible for the current through a QPC between two superfluids of strongly interacting Fermi gas under spin-dependent particle loss. We observed no critical behavior, which contrasts with other types of superfluid-suppressing perturbations such as moving defects \cite{madison_vortex_2000, sobirey_observation_2021, del_pace_imprinting_2022}, magnetic fields in superconductors \cite{scheer_conduction_2000}, and photon absorption by superconducting nanowire single-photon detectors \cite{natarajan_superconducting_2012}. While our most significant observations are captured by our model, deviations from the theory point to the importance of strong interactions in the unitary regime and energy dependence of the channel and dissipation. These effects can be probed in future studies on thermoelectricity \cite{husmannBreakdownWiedemannFranz2018, hauslerInteractionAssistedReversalThermopower2021}, spin conductance \cite{krinner_mapping_2016, visuri_giamarchi_2020}, and correlated loss \cite{partridge_molecular_2005, werner_number_2009, paintner_pair_2019, liu_observation_2021} in similar systems.

\section*{Acknowledgments}
We thank Alexander Frank for technical support and Lukas Rammelm\"uller for providing us with his calculations of the finite-temperature equation of state of the spin-polarized unitary Fermi gas. We also thank Giulia Del Pace, Eugene Demler, Alex G\'omez Salvador, Mart\'on Kan\'asz-Nagy, and Alfredo Levy Yeyati for discussions. M.-Z.H., J.M., P.F., M.T., S.W., and T.E. acknowledge the Swiss National Science Foundation (Grants No.~182650, No.~212168 and No.~NCCR-QSIT) and European Research Council advanced grant TransQ (Grant No.~742579) for funding. A.-M.V. acknowledges funding from the Deutsche Forschungsgemeinschaft (DFG, German Research Foundation) in particular under project no.~277625399 - TRR 185 (B3) and project no.~277146847 - CRC 1238 (C05) and Germany’s Excellence Strategy -- Cluster of Excellence Matter and Light for Quantum Computing (ML4Q) EXC2004/1 -- 390534769. S.U. acknowledges MEXT Leading Initiative for Excellent Young Researchers, JSPS KAKENHI (Grant No.~JP21K03436), and Matsuo Foundation. T.G. acknowledges support from the Swiss National Science Foundation (Grant No.~2000020-188687).

M.-Z. H. and J. M. contributed equally to this work.

\bibliography{paper}

\clearpage
\section*{Supplemental material}
\setcounter{figure}{0}
\renewcommand{\thefigure}{S\arabic{figure}}

\section{Summary of the theoretical calculation}
\label{sec:theory}

\subsection{Lossy quantum dot coupled to reservoirs}

We model the channel between the reservoirs by a lossy quantum dot coupled to leads, described as an open quantum system for which the dynamics of the density operator $\hat{\rho}$ obeys the Lindblad master equation~\cite{breuer2002}
\begin{equation}
    \dv{\hat{\rho}}{t} = -i [\hat{H}, \hat{\rho}] + \sum_{\sigma = \uparrow, \downarrow} \gamma_{\sigma} \left( \hat{d}_{\sigma}^{\phantom{\dagger}}\hat{\rho} \hat{d}_{\sigma}^{\dagger} - \frac{1}{2} \left\{ \hat{d}_{\sigma}^{\dagger} \hat{d}_{\sigma}^{\phantom{\dagger}}, \hat{\rho} \right\} \right).
\label{eq:master_equation_quantum_dot}
\end{equation}
Here, $\gamma_{\sigma}$ is the dissipation rate of spin state $\sigma$ and $\hat{d}_{\sigma}^{\phantom{\dagger}}$ ($\hat{d}_{\sigma}^{\dagger}$) the fermionic annihilation (creation) operator at the quantum dot.
Notice also that we adopt natural units $\hbar=k_B=1$ \newnew{in this section}. 
The loss within the channel is taken into account by a local particle loss at the quantum dot. The Hamiltonian is $\hat{H} = \hat{H}_L + \hat{H}_R + \hat{H}_d + \hat{H}_t$, where $\hat{H}_L$ and $\hat{H}_R$ describe the superfluid reservoirs, $\hat{H}_d$ the quantum dot, and $\hat{H}_t$ the tunneling between the reservoirs and the dot.

We treat effects of the superfluid reservoirs within the mean-field Hamiltonian (see Ref.~\cite{Husmann_quantum_point_contact2015})
\begin{equation}
    \hat{H}_i = \sum_{\vec{k}} \hat{\Psi}_{i \vec{k}}^{\dagger}\left[ \left( \epsilon_{\vec{k}} - \mu_i \right) \sigma_z + \Delta \sigma_x \right] \hat{\Psi}_{i \vec{k}}
\label{eq:reservoir_hamiltonian}
\end{equation}
where $i = L, R$ for left and right and $\epsilon_{\vec{k}}$ represents the single-particle energy at momentum $\vec{k}$. As in the case of Ref.~\cite{Husmann_quantum_point_contact2015}, we consider a linear single-particle dispersion relation $\epsilon_{\vec{k}} = \hbar v_F(k - k_F)$, corresponding to a constant density of states in the normal state (cf.~Ref.~\cite{setiawan_analytic_2022}) as opposed to the quadratic density of states for harmonically-trapped gases. The chemical potential in each reservoir is $\mu_i$, $\Delta$ is the superfluid energy gap, and $\sigma_{x, z}$ are Pauli spin matrices. Here, $\hat{\Psi}_{i \vec{k}}$ denotes the Nambu spinor $\hat{\Psi}_{i \vec{k}} = \left(\hat{\psi}_{i \vec{k} \uparrow}, \: \hat{\psi}_{i -\vec{k} \downarrow}^{\dagger} \right)^T$, and $\hat{\psi}_{i \vec{k} \sigma}^{\dagger}$ ($\hat{\psi}_{i \vec{k} \sigma}$) is the fermionic creation (annihilation) operator for reservoir~$i$.
A current through the quantum dot is induced by a chemical potential difference $\Delta\mu = \mu_L - \mu_R$ between the two reservoirs. We choose the chemical potentials symmetrically as $\mu_L = \Delta\mu/2$ and $\mu_R = -\Delta\mu/2$. While the reservoirs are interacting, we consider noninteracting fermions at the quantum dot, with the Hamiltonian $\hat{H}_d = \epsilon_d \sum_{\sigma = \uparrow, \downarrow} \hat{d}_{\sigma}^{\dagger} \hat{d}_{\sigma}$.
Here, $\epsilon_d$ is the quantum dot energy level. Single-particle tunneling occurs between the position $\mathbf{r} = \mathbf{0}$ in each reservoir and the quantum dot:
\begin{align}
    \hat{H}_t &= -\tau \sum_{\sigma = \uparrow, \downarrow} \left[\hat{\psi}^{\dagger}_{L\sigma}(\mathbf{0}) \hat{d}_{\sigma} + \hat{d}_{\sigma}^{\dagger} \hat{\psi}^{\phantom{\dagger}}_{R\sigma}(\mathbf{0}) + \text{h.c.}\right].
\label{eq:hamiltonian_tun_qd}
\end{align}
The tunneling amplitude $\tau$ is energy-independent and is a fitted parameter as are the dissipation rates $\gamma_{\sigma}$. Note that $\tau$ has units of \si{(energy) \cdot (length)^{3/2}} for 3D reservoirs. Its corresponding energy scale, the linewidth or inverse lifetime of the quantum dot, is given by $\Gamma_d = \pi\rho_0\tau^2$ where $\rho_0$ is the density of state of the reservoirs at the Fermi level in the normal state. The quantum dot energy level is set to $\epsilon_d = 0$ in the middle of the reservoir chemical potentials, where the high transparency is realized~\cite{martin2011josephson}.

Transport is connected to the change in particle numbers of the reservoirs $\hat{N}_i = \hat{N}_{i \uparrow} + \hat{N}_{i \downarrow}$, where $\hat{N}_{i \sigma} = \int d\mathbf{r} \hat{\psi}_{i \sigma}^{\dagger}(\mathbf{r}) \hat{\psi}_{i \sigma}^{\phantom{\dagger}}(\mathbf{r})$. For an open quantum system described by the quantum master equation~(\ref{eq:master_equation_quantum_dot}), the time derivative of the particle number is obtained as (see Ref.~\cite{visuri2022})
\begin{align}
    \dv{t} \braket{\hat{N}_i} &= \dv{t} \text{Tr}\left( \hat{N}_i \hat{\rho}(t) \right) 
    \label{eq:trace} \\
    &= i \tau \sum_{\sigma = \uparrow, \downarrow} \left( \braket{\psi^{\dagger}_{i\sigma}(\mathbf{0}) d_{\sigma}^{\phantom{\dagger}}} - \braket{d_{\sigma}^{\dagger} \psi_{i\sigma}^{\phantom{\dagger}}(\mathbf{0})} \right).
\end{align}
The apparent current observed in the experiment is given by $I_N = -\frac{1}{2} \dv{t}\braket{\hat{N}_L - \hat{N}_R}$ and the loss rate is $\dot{N} = \dv{t}\braket{\hat{N}_L + \hat{N}_R} = -\sum_{\sigma} \gamma_{\sigma} \braket{\hat{d}_{\sigma}^{\dagger} \hat{d}_{\sigma}^{\phantom{\dagger}}}$.

\subsection{Keldysh formalism}

We apply the Keldysh formalism~\cite{kamenev2011, sieberer2016} to compute the nonequilibrium expectation values, which can be expressed in terms of Keldysh Green's functions. By using the path integral formulation, the Keldysh action is written in the basis of fermionic coherent states parametrized by the Grassmann variables $\psi$. Integration over a closed time contour is performed by introducing $\psi = (\psi^+, \psi^-)$ for the forward and backward time branches. For the convenience of calculation, we perform  the following change of the field variables~\cite{kamenev2011}:
\begin{equation}
    \begin{pmatrix}
        \psi^1\\
        \psi^2
    \end{pmatrix}    
    =\frac{1}{\sqrt{2}}\begin{pmatrix}
        1 & 1\\
        1 & -1
    \end{pmatrix}
    \begin{pmatrix}
        \psi^+\\
        \psi^-
    \end{pmatrix},
\end{equation}
which allows to express the expectation values in terms of advanced, retarded, and Keldysh components of Green's function. As we are interested in equal-time correlations, it is also convenient to use the frequency basis. The relevant expectation values only depend on $\psi$ and $\bar{\psi}$ at $\mathbf{r} = \mathbf{0}$, and in the following, we denote $\psi(\omega) = \psi(\mathbf{r} = \mathbf{0}, \omega)$. Two-operator expectation values are calculated with Gaussian integration of the Grassmann variables
\begin{equation}
    \braket{\psi^a \bar{\psi}^b} = \int \mathcal{D}[\bar{\psi}, \psi] \psi^a \bar{\psi}^{b} e^{i S[\bar{\psi}, \psi]} = i \mathcal{G}_{a b},
\label{eq:correlation_function}
\end{equation}
where $a, b$ are sets of relevant indices ($i$, $\sigma$, 1, 2) and $\mathcal{G}_{a b}$ denotes the matrix element of the Green's function. The current is given by
\begin{align}
    I_N = \frac{i \tau}{4} \sum_{\sigma = \uparrow, \downarrow} \int \frac{d \omega}{2 \pi} \big( \braket{d_{\sigma}^1 \bar{\psi}_{L \sigma}^1} - \braket{\psi_{L \sigma}^1 \bar{d}_{\sigma}^1} \nonumber \\
    + \braket{\psi_{R \sigma}^1 \bar{d}_{\sigma}^1} 
    - \braket{d_{\sigma}^1 \bar{\psi}_{R \sigma}^1} \big)
\end{align}
and the particle density at the quantum dot by 
\begin{equation}
    \braket{n_{d \sigma}} = \frac{1}{2} \sum_{\sigma = \uparrow, \downarrow} \int \frac{d \omega}{2 \pi} \left( \braket{d_{\sigma}^1 \bar{d}_{\sigma}^1} - \braket{d_{\sigma}^1 \bar{d}_{\sigma}^2} + \braket{d_{\sigma}^2 \bar{d}_{\sigma}^1} \right).
\end{equation}

The action $S = \int \frac{d \omega}{2 \pi} \bar{\mathbf{\Psi}}(\omega) \mathcal{G}^{-1}(\omega) \mathbf{\Psi}(\omega)$ can be written as the sum~\cite{sieberer2016}
\begin{equation}
    S = S_L + S_R + S_d + S_t + S_{\text{loss}},
    \label{eq:action_qd}
\end{equation}
where the first four terms arise from the Hamiltonian and the last one from the loss term in Eq.~(\ref{eq:master_equation_quantum_dot}). For uncoupled reservoirs, $S_{L, R}$ is written in terms of the four-component Nambu-Keldysh spinor $\mathbf{\Psi}_i = \left(\psi_{i \uparrow}^1, \bar{\psi}_{i \downarrow}^1,	\psi_{i \uparrow}^2, \bar{\psi}_{i \downarrow}^2 \right)^T$ and inverse Green's function $\mathcal{G}_i^{-1}$ has the structure
\begin{equation}
    \mathcal{G}_i^{-1} = 
    \begin{pmatrix}
    0	&\left[ g_i^A \right]^{-1} \\
    \left[ g_i^R \right]^{-1}	&[G_i^{-1}]^{K}
\end{pmatrix},
\label{eq:general_inverse_greens_function}
\end{equation}
where $[G_i^{-1}]^{K} = -\left[ g_i^R \right]^{-1} \left[ g_i^K \right]\left[ g_i^A \right]^{-1}$, and $g_i^A$, $g_i^R$, and $g_i^K$ are the advanced, retarded, and Keldysh components of the Green's function.
For conventional $s$-wave Fermi superfluids, $g_i^A$, $g_i^R$, and $g_i^K$ are expressed by matrices of size $2 \times 2$. The functions $[g_i^A]^{-1}$ and $[g_i^R]^{-1}$ are obtained locally at $\mathbf{r} = \mathbf{0}$ as~\cite{cuevas_hamiltonian_1996, Husmann_quantum_point_contact2015}
\begin{align}
    [g_i^{R, A}]^{-1} = \frac{W}{\sqrt{\Delta^2 - (\bar{\omega} \pm i\eta)^2}} 
    \begin{pmatrix}
    \bar{\omega} \pm i\eta	&\Delta	\\
    \Delta					&\bar{\omega} \pm i\eta
    \end{pmatrix},
    \label{eq:local_greens_functions}
\end{align}
where \new{$W=1/(\pi\rho_0)$} and $\eta > 0$ is an infinitesimal constant which regularizes the Green's functions. The upper (lower) sign corresponds to the retarded (advanced) Green's function. We define the frequency relative to the chemical potential as $\bar{\omega} = \omega - \mu_i$. By assuming that the reservoirs are in a thermal state at all times, the Keldysh component $g^K$ can be obtained through the fluctuation-dissipation theorem, $g^K = (g^R - g^A)[1 - 2 n_F(\bar{\omega})]$. Here, $n_F(\omega)= \left(e^{\omega/T} + 1\right)^{-1} $ with temperature $T$ denotes the Fermi-Dirac distribution. The inverse Green's function for the non-interacting quantum dot contains the dissipation term of Eq.~(\ref{eq:master_equation_quantum_dot}) (see Refs.~\cite{jinGenericTransportFormula2020, visuri2022, uchino_comparative_2022}). It has the same form as Eq.~(\ref{eq:general_inverse_greens_function}) with the different components given by
\begin{align}
    &[g_d^{R, A}(\omega)]^{-1} = 
    \begin{pmatrix}
        \omega - \epsilon_d \pm \frac{i \gamma_{\uparrow}}{2}	&0 \\
        0	&-(\omega - \epsilon_d) \pm \frac{i \gamma_{\downarrow}}{2}
    \end{pmatrix}, 
    \label{eq:quantum_dot_retarded_advanced} \\
    &[G_d^{-1}(\omega)]^{K} = 
    \begin{pmatrix}
        i \gamma_{\uparrow}	&0 \\
        0	&- i \gamma_{\downarrow}
    \end{pmatrix}.
\label{eq:quantum_dot_Keldysh_component}
\end{align}

In the absence of a bias, the action including the left and right reservoirs, tunneling, and loss could be represented by an $8 \times 8$ matrix $\mathcal{G}^{-1}$. One must take into account that in the presence of $\Delta$, fermions in either reservoir at equal but opposite frequencies $\pm \bar{\omega}$ relative to the chemical potential are coupled, while the tunneling term couples fermions on the left and right sides at the same absolute frequency $\omega$. Therefore, a finite bias leads to an action which is not block diagonal in frequency, and is represented by an infinite-size matrix with elements at increasing discrete frequencies. This corresponds physically to multiple Andreev reflections. As the off-diagonal matrix elements $[g^{A, R}]_{12}^{-1}$, $[g^{A, R}]_{21}^{-1}$ decay with $\bar{\omega}$, the matrix can be truncated at a certain order of the tunneling process according to desired accuracy. It can then be inverted numerically to obtain the matrix elements $\mathcal{G}_{a b}$.

A spin bias $\Delta b = (\Delta\mu_\downarrow - \Delta\mu_\uparrow)/2$ between the reservoirs may be modeled by including a magnetic field term in the local Green's functions \cite{lu_spin-polarized_2020}, so that Eq.~(\ref{eq:local_greens_functions}) is replaced by
\begin{equation}
    [g^{R, A}]^{-1} = 
    \begin{pmatrix}
        \frac{W(\bar{\omega} + b \pm i\eta)}{\sqrt{\Delta^2 - (\bar{\omega} + b \pm i\eta)^2}}	&\frac{W\Delta}{\sqrt{\Delta^2 - (\bar{\omega} + b \pm i\eta)^2}}	\\
        \frac{W\Delta}{\sqrt{\Delta^2 - (\bar{\omega} + b \pm i\eta)^2}}					&\frac{W(\bar{\omega} - b \pm i\eta)}{\sqrt{\Delta^2 - (\bar{\omega} - b \pm i\eta)^2}}
    \end{pmatrix}.
\end{equation}
Here, we have defined $\mu_i = (\mu_{i \uparrow} + \mu_{i \downarrow})/2$ and $b_i = (\mu_{i \uparrow} - \mu_{i \downarrow})/2$. This minimal approach corresponds to pairing with an energy offset from the Fermi level on either side, and it does not take into account the spatially modulated order parameter which typically arises from FFLO pairing at a finite center-of-mass momentum.


For numerical calculation of the observables, the frequency integral must be truncated to a range $\bar{\omega} \in [-\Lambda,\Lambda]$ given by a high energy cutoff $\Lambda$. The integrand of the current $I_N$ falls off quickly for $\abs{\bar{\omega}} > \Delta$ so $I_N$ is independent of $\Lambda$. 
Note that the occupation of the quantum dot is bounded due to its Lorentzian lineshape \newnew{in frequency}, though it contains contributions from an unphysically large range of energies since the fitted linewidth is several times the gap. 

\subsection{\new{Effects of dissipation}}

\new{As seen in Fig.~2b of the main text, the current shows a transition from a highly nonlinear current-bias relation to a linear one as the dissipation rate increases to $\hbar\gamma>\Delta$. The calculated current due to MAR, which fits well the measured current, is still significantly higher than the current expected from reservoirs in the normal state. The transition from a nonlinear to linear current-bias relation can be understood to result from two mechanisms: 1) dissipation suppresses more strongly the higher-order MAR processes, which are the main contributor to the current at lower bias; 2) dissipation also effectively broadens the density-of-state of the reservoirs,
allowing lower-order MAR (fewer pairs co-tunneling) to contribute at a given bias. This is confirmed by our numerical observation that for increasing $\gamma_\sigma$, the calculation of the current converges at smaller sizes of the inverse Green’s function matrix $\mathcal{G}^{-1}(\omega)$.
The size of this matrix represents the maximum order of MAR that is accounted for.}

\new{It is worth mentioning a similar transition caused by quasiparticle dephasing due to inelastic scattering studied in superconducting QPCs \cite{cuevas_hamiltonian_1996, martin-rodero_microscopic_1996}: higher-order MAR are more strongly damped, rendering the nonlinear current a linear conductance $G\sim\Delta/\eta h$ in the regime $\Delta\mu<\eta$ where $\eta$ is the damping rate showed up in Eq.~12. In our case, the particle dissipation takes a similar role as the quasiparticle dephasing.} 

\section{Comparison to other transport mechanisms between superfluids}

\new{There are many types of both reversible and irreversible processes that can occur when connecting two superfluids with a weak link, all depending on the length scales involved and the boundary conditions imposed by the superfluid reservoirs \cite{tinkham_introduction_1996}, including Josephson currents, MAR, phase slips, and excitations of gapless sound modes. It can be shown very generally in a mean field approximation that the current through such weak links \cite{likharev_superconducting_1979} induced by a chemical potential bias is \cite{cuevas_hamiltonian_1996}
\begin{align*}
    &I(t) = \sum_{n=-\infty}^\infty I_n(\Delta\mu) e^{-i n \omega_J t} 
    = I_0(\Delta\mu) + \\ & + 2\sum_{n=1}^\infty \mathrm{Im}[I_n(\Delta\mu)] \sin(n\omega_J t) + \mathrm{Re}[I_n(\Delta\mu)] \cos(n\omega_J t)
\end{align*}
where $\omega_J = \Delta\mu/\hbar$ is the Josephson frequency. The Fourier components $I_n$, which originate from tunneling processes of $n$ Cooper pairs, depend on the bias $\Delta\mu$ and the transparency of the contact $\alpha$. The dc component $I_0$ is an irreversible current carried via MAR since it leads to Joule heating $I_0\Delta\mu$ while the ac components are the reversible Josephson currents that generate no heat. Our system, composed of two finite-size superfluid reservoirs connected by a ballistic channel, is therefore conceptually similar to an ideal Josephson junction carrying the reversible currents shunted by a capacitor (the reservoirs) and a nonlinear resistor carrying the irreversible current (a model widely used e.g.~in \cite{tinkham_introduction_1996,del_pace_tunneling_2021}). The capacitive and inductive Josephson energy contained in the system can therefore be dissipated by the irreversible currents which suppresses the reversible dynamics \cite{meier_josephson_2001}.}

\new{At biases $\Delta\mu$ small compared to the superfluid gap $\Delta$, the irreversible components decay quickly with decreasing $\alpha$ as they require high-order MAR processes $n_\mathrm{pair}\sim\Delta/\Delta\mu$ which succeed with probability $\sim \alpha^{n_\mathrm{pair}}$. Meanwhile, the reversible components decay slowly, giving rise to the well-known expression for the ac Josephson current $I = I_c \sin(\omega_J t)$ in the limit $\Delta\mu\ll(1-\alpha)\Delta$. The observation of the Josephson oscillation in cold-atom systems is achieved with a wide tunnel junction where many transport modes are available $\nch\gg1$ but they all have low transparency $\alpha\ll1$ so that there is effectively no irreversible current to dissipate the system's energy \cite{valtolina_roati2015}. Our experiment, on the other hand, takes place in the exact opposite limit where $\nch\gtrsim1$ and $\alpha\approx1$. The contact has near-perfect transparency---confirmed by the observation of conductance quantized to multiples of $1/h$ for weak interactions \cite{krinnerObservationQuantizedConductance2014}---since any imperfection requires structures smaller than the Fermi wavelength which is below the diffraction limit in our case. In the ballistic regime $\alpha\approx1$, the irreversible and reversible currents are comparable and on the order of $\Delta/h$ \cite{cuevas_hamiltonian_1996}, so the irreversible current can efficiently damp reversible Josephson oscillations to levels below our experimental resolution \cite{yao_controlled_2018}. Furthermore, the reversible currents are exponentially suppressed at finite bias $\Delta\mu>(1-\alpha)\Delta$, where the irreversible current from MAR prevails. These characteristics are indeed compatible with our observations. 
}

\new{On the other hand, gapless sound modes in the superfluids (Anderson-Bogoliubov/Nambu-Goldstone modes in the BCS regime or Bogoliubov quasiparticles in the BEC regime) can allow coherent pair-tunneling to contribute to irreversible currents, as studied in e.g. Ref.~\cite{del_pace_tunneling_2021}. Recent work \cite{uchino_bosonic_2020, uchino_role_2020} has theoretically studied the role of these sound modes in a QPC in the BEC and BCS limits. Although the continuation to the unitary limit is unclear, the contribution to the conductance from these modes is expected to be on the order of $1/h$ per mode. In the case of a QPC with low transparency or the tunnel junction limit, as in Ref.~\cite{del_pace_tunneling_2021}, the conductance originates from these modes and can indeed outweigh the MAR current. This is a primary difference between our system and the 2D tunnel junction explored in Ref.~\cite{del_pace_tunneling_2021}, where the tunneling probability is $\sim 10^{-2}$ but $10^4$ modes contribute, leading to a normal conductance of $\sim10^2/h$. Thus the sound-mode conductance of $\sim 1/h$ per mode dominates and reaches 100 times the quasiparticle conductance. In our case, there are only a few modes, so we expect the Goldstone mode contribution to be negligible compared to the MAR current.}


\section{Experimental details}

\subsection{Experimental cycle}
We begin by preparing a non-degenerate, spin-balanced cloud with approximately \num{2e5} atoms in the first and third-lowest spin states of \textsuperscript{6}Li ($\ket{\downarrow}$ and $\ket{\uparrow}$ respectively) at their Feshbach resonance $B=\SI{689.7}{G}$. The cloud is trapped along $y$ (direction of transport) by a magnetic trap and along $x$ and $z$ by an optical dipole trap propagating along $y$. We then prepare the initial imbalance $\Delta N$ by shifting the cloud along $y$ with a magnetic field gradient, then ramping up the power of the beams that define the QPC along with a repulsive ``wall'' beam to block transport (Sec.~\ref{sec:defining_the_quantum_point_contact}), and finally returning the cloud center to coincide with the QPC. We then evaporatively cool the cloud to degeneracy by ramping down the dipole trap power with a magnetic field gradient along gravity to further lower the trap depth. At this stage, just before transport, the trap frequencies are $\nu_{x,\mathrm{trap}} = \SI{184.7(6)}{Hz}$, $\nu_{y,\mathrm{trap}} = \SI{28.2(1)}{Hz}$, $\nu_{z,\mathrm{trap}} = \SI{178.4(5)}{Hz}$ and the cloud has $N=\num{195(14)e3}$ atoms at $T=\SI{100(2)}{nK}$, corresponding to $T/T_F=0.26(1)$, and an atom number imbalance of $\Delta N/N = 0.37(1)$. There is a very small spin imbalance $\Delta M = M_L - M_R \lesssim 0.02 N$ that is due to a small overall imbalance in the populations of the two spin states $M = N_\downarrow - N_\uparrow \lesssim 0.02 N$.

Because we prepare the density imbalance before evaporation, the evaporation efficiencies of the two reservoirs differ due to the differing atom number in each one. This leads to a large initial temperature bias between the two reservoirs $\Delta T$ which gives rise to a strong, unwanted thermoelectric effect \cite{husmannBreakdownWiedemannFranz2018}. To suppress this effect, we apply a magnetic field gradient during evaporation to compress the reservoir with lower atom number and decompress the one with more atoms, thereby equalizing their evaporation efficiencies such that $\Delta T \approx 0$ at the beginning of transport.

We then ramp up the powers of the attractive gate and dissipation beams. Then transport is allowed by switching off the wall beam before switching it on at a later time to block transport again. All beam powers except the dipole trap and the wall are ramped down such that each reservoir is in a half-harmonic trap, at which point we take an absorption image of both spin states by sending two pulses of light resonant with each spin state in quick succession (\SI{225}{\micro s} apart) synchronized to a CCD camera operating in fast kinetics acquisition mode.

\subsection{Defining the quantum point contact}
\label{sec:defining_the_quantum_point_contact}

Two TEM\textsubscript{01}-like beams of repulsive 532 nm light define the QPC at the intersection of their nodal planes. The first propagates along $x$, has a Gaussian waist $w_{z,\mathrm{QPC}}=\SI{30.2}{\micro m}$ along $y$, and has a peak confinement frequency of $\nu_{z,\mathrm{QPC}}=\SI{9.9(2)}{kHz}$ along $z$. The second propagates along $z$, has a Gaussian waist of $w_{x,\mathrm{QPC}}=\SI{6.82}{\micro m}$ along $y$, and a peak confinement frequency of \new{$\nu_{x,\mathrm{QPC}}=\SI{10(2)}{kHz}$} along $x$ for the data presented. The light intensity is not exactly zero at these beams' nodal planes which leads to an additional contribution to their zero-point energy that is non-negligible. This is calibrated by measuring the onset of quantized conductance in a non-interacting system ($B=\SI{568}{G}$) under the same conditions \cite{krinnerObservationQuantizedConductance2014}. The wall beam is a 532 nm elliptical beam that also propagates along $z$ and has a waist of \SI{8.58}{\micro m} along $y$.

The Gaussian gate beam of attractive 777\,nm light propagates along $z$ and has waists of $w_{x,\mathrm{gate}}=\SI{30.3}{\micro m}$ along $x$ and $w_{y,\mathrm{gate}}=\SI{31.7}{\micro m}$ along $y$. The potential energy it exerts on the atoms $V_g<0$ was determined by calibrating its power at the location of the atoms and computing the resulting ac-Stark shift from its known beam profile and wavelength. As this beam is essentially a weakly-focused optical tweezer, it has a transverse confinement frequency $\nu_{x,\mathrm{gate}} = \sqrt{\abs{V_g}/m}/\pi w_{x,\mathrm{gate}} = \SI{220(2)}{Hz}$ which adds in quadrature to $\nu_{x,\mathrm{QPC}}$ to determine the net confinement frequency at the center of the QPC. We observed that this power calibration drifted over the course of our measurements. As a result, there is a relatively large systematic uncertainty on this gate potential $V_g$ of approximately 6\%. The equilibrium imbalance $\Delta N$ is extremely sensitive to the center position of this beam along $y$, especially in the presence of dissipation, on a scale of $\sim\SI{1}{\micro m}$---many times smaller than the beam's waist. We therefore align this beam position such that, starting from zero imbalance $\Delta N=0$, the imbalance remains zero at long time for each dissipation strength.

\subsection{Dissipation mechanism}
\label{sec:dissipation_mechanism}

The dissipative beam is a tightly-focused Gaussian beam propagating along $z$ aligned to the center of the QPC with waists $w_{x,\mathrm{dis}} = \SI{1.28}{\micro m}$ and $w_{y,\mathrm{dis}} = \SI{1.31}{\micro m}$. It is shaped and positioned using a digital micromirror device whose Fourier plane is projected onto the plane of the QPC by a high-NA microscope objective \cite{lebratQuantizedConductanceSpinSelective2019}. Its frequency is tuned into resonance with the $\sigma^+$-transition between $\kdn = \ket{-1/2,1} - \epsilon\ket{1/2,0}$ and $\ket{e} \approx \ket{3/2,0}$ (written in the basis $\ket{m_J,m_I}$), 2.98\,GHz blue-detuned from the center of the $D_2$ line \cite{corman_quantized_2019}. The excited state $\ket{e}$ has an inverse lifetime of $\Gamma_e/2\pi=\SI{5.87}{MHz}$ and decays into the first $\ket{1}=\kdn$, fifth $\ket{5}=\ket{1/2,0} + \epsilon\ket{-1/2,1}$, and sixth $\ket{6}=\ket{1/2,1}$ lowest ground states with branching ratios $\Gamma_{e1}=\num{2.9e-3}\Gamma_e$, $\Gamma_{e5}=0.997\Gamma_e$, and $\Gamma_{e6}=\num{8e-7}\Gamma_e$. This ensures that only 0.29\% of $\kdn$ atoms that have scattered a photon return to $\kdn$ and the rest are optically pumped to $\ket{5}$. These states are then out of resonance, anti-trapped by the magnetic field as they are low field-seeking, and have a recoil energy larger than the optical trap depth so they are quickly lost from the system. We therefore interpret a photon scattering event by a $\kdn$ atom as a quantum jump of the loss process modeled in Sec.~\ref{sec:theory}. We use this optical pumping process instead of photon scattering on a closed transition since the strongly-interacting $\kdn$ atoms imparted with the photon recoil energy quickly destroy the cloud.

We calibrate the power $P_\mathrm{dis}$ of this beam at the location of the atoms and use the known beam shape to compute its peak intensity $I_\mathrm{dis}=2P_\mathrm{dis}/(\pi w_{x,\mathrm{dis}} w_{y,\mathrm{dis}})$, from which we can compute the photon scattering rates $\Gamma_\sigma$ and ac-Stark shifts $V_\sigma$ of $\kdn$ and $\kup$ \cite{corman_quantized_2019} which are directly proportional to $I_\mathrm{dis}$ since all intensities used in our measurements are less than \new{0.5\%} of the saturation intensity of this transition $I_\mathrm{sat}=\SI{8.7}{kW/m^2}$. In addition to the intensity and frequency, it is also crucial to know the polarization of the beam to compute the photon scattering rates and ac-Stark shifts. We measured the fraction of the total power in each polarization directly on the atoms by monitoring atom loss rate in a non-interacting system. We tuned the beam's frequency into resonance with the $\sigma^-$, $\pi$, and $\sigma^+$ transitions of $\kdn$'s $m_J=-1/2$ component and measured the atom loss rate at constant power by fitting an exponential to $N_\downarrow(t)$. The ratio between each loss rate scaled by each transition's oscillator strength is then the ratio of the powers in each polarization. This calibration yielded 78.7(5)\% of the total power in $\sigma^+$, 6.6(1)\% in $\pi$, and 14.8(6)\% in $\sigma^-$. Given these calibrations and $\Gamma_\downarrow$, we have $\Gamma_\uparrow=\num{5.4e-4}\Gamma_\downarrow$, $V_\uparrow=k_B\times(\SI{1.49}{nK/kHz})\Gamma_\downarrow$, and $V_\downarrow=k_B\times(\SI{1.37}{nK/kHz})\Gamma_\downarrow$ which are all negligible compared to other relevant scales.

In the non-interacting gas ($B=\SI{568}{G}$), we observe that photon scattering of $\kdn$ does not induce any loss of $\kup$---strong evidence for the interactions between $\kup=\ket{3}$ and $\ket{5}$ being weak as expected \cite{houbiers_elastic_1998}, however at unitarity ($B=\SI{689.7}{G}$), there are significant losses of $\kup$ and its loss rate is proportional to the loss rate of $\kdn$ $\dot{N}_\uparrow \approx 0.7 \dot{N}_\downarrow$. This indicates that the loss is due to the $s$-wave interactions between $\kdn$ and $\kup$. This is supported by the observation that the relative loss ratio $r=\dot{N}_\uparrow/\dot{N}_\downarrow$ decreases with increasing temperature and decreasing chemical potential where interatomic binding is weaker. 
\new{For a given QPC configuration, we observe a rather robust loss ratio $r$, allowing us to simplify the fit by fixing $\gamma_\uparrow$ to $r\gamma_\downarrow$ with $r$ determined from the data. (See Sec.~\ref{sec:fitting_procedure} for details)}

\new{Microscopically, when a $\ket{\downarrow}$ atom absorbs a photon, energy can be transferred to $\ket{\uparrow}$ atoms via two mechanisms that both originate from contact interactions: 1) part of the recoil energy absorbed by $\ket{\downarrow}$ is transferred to $\ket{\uparrow}$ via interactions and 2) the conversion from $\ket{\downarrow}$ to $\ket{e}$ changes the interaction strength ($s$-wave scattering length) with $\ket{\uparrow}$, thereby depositing energy via Tan's dynamic sweep theorem \cite{zwergerBCSBECCrossoverUnitary2012}. Photon absorption by $\ket{\downarrow}$ atoms can therefore promote $\ket{\downarrow}$ atoms to high momentum states which, if their kinetic energy is larger than the trap depth, can cause them to escape the trap and be effectively dissipated. 
These unpaired $\kup$ atoms will acquire momentum predominantly perpendicular to the 2D region around the QPC. The strong attractive gate beam also provides an easier escape path within the 2D region, limiting the chance of their depositing energy to the reservoirs. This picture is compatible with the observation.}

\section{Reservoir thermodynamics}
\label{sec:reservoir_thermodynamics}

For each shot, we obtain the column density $n_{i\sigma}^\mathrm{col}(y,z)$ of both half-harmonic reservoirs $i=L,R$ and both spin states $\sigma=\downarrow,\uparrow$ from an absorption image taken \textit{in situ} along the $x$-direction with a calibrated imaging system. The atom number in each spin state and reservoir is determined by integrating the column density over the half-plane of each reservoir
\begin{align}\begin{split}
    N_{L\sigma} &= \int_{-\infty}^0 \dd{y} \int_{-\infty}^\infty \dd{z} n_{L\sigma}^\mathrm{col}(y,z) \\
    N_{R\sigma} &= \int_0^\infty \dd{y} \int_{-\infty}^\infty \dd{z} n_{R\sigma}^\mathrm{col}(y,z).
\end{split}\end{align}
We then compute the second spatial moment of the line density $n_{i\sigma}^\mathrm{lin}(y) = \int_{-\infty}^\infty \dd{z} n_{i\sigma}^\mathrm{col}(y,z)$
\begin{equation}
    \ev{y^2}_{i\sigma} = \frac{\int_{-\infty}^\infty \dd{y} n_{i\sigma}^\mathrm{lin}(y) y^2}{\int_{-\infty}^\infty \dd{y} n_{i\sigma}^\mathrm{lin}(y)}
\end{equation}
which determines the total energy per particle through the virial theorem for the spin-balanced, harmonically-trapped unitary fermi gas $E_{i\sigma}/N_{i\sigma} = 3m\omega_y^2 \ev{y^2}_{i\sigma}$ \cite{thomas_virial_2005}. With the $N_{i\sigma}$ and $E_{i\sigma}$ along with the average trap frequency $\bar{\omega} = (\omega_x\omega_y\omega_z)^{1/3}$, we use the equation of state (EoS) to solve for the temperature $T_{i\sigma}$.
\newnew{We note, however, that our trap along $x$ and $z$ by the optical dipole trap is not perfectly harmonic due to the beam's gaussian profile. The trap anharmonicity becomes non-negligible when the trap depth is low such that the cloud size is not too small compared to the beam waist. In our case, we use low trap depth to achieve a low temperature so the real thermodynamic quantities (such as chemical potential and temperature) slightly deviate from those derived from the virial theorem assuming a harmonic potential. Moreover, the channel-defining potentials also modify the reservoir EoS during transport, different from the trap for thermometry. Nevertheless, we estimate from a simulation using all calibrated potentials (Sec.~\ref{sec:defining_the_quantum_point_contact}) that the true $\Delta\mu$ can be up to $15\%$ larger than the value determined by assuming harmonic potentials. Given the relatively large uncertainty in the estimated superfluid gap, which limits the uncertainty in $\Delta\mu/\Delta$ which is most relevant for this study, we neglect the effect of trap anharmonicity and that of the channel-defining beams in this work.}

Since the dissipation rate is spin-dependent, a spin polarization builds up over time and the virial theorem is no longer valid. However, since the magnetization in each reservoir is small $M/N < 10\%$ for most of the data, the shape of the minority spin cloud do not change significantly \cite{shinPhaseDiagramTwocomponent2008, olsen_phase_diagram_2015} and we can apply the same temperature extraction procedure with limited systematic error. This approach is validated by the observation that the extracted temperature does not depend on dissipation strength or transport time within statistical measurement fluctuations. This observation also indicates that thermoelectric effects are negligible in our parameter regime (e.g.~density or spin currents coupling to entropy currents that lead to a temperature bias).

With the measured atom number in both spin states and their temperature in each reservoir $N_{i\downarrow}, N_{i\uparrow}, T_i$, we use the recently-computed EoS of the spin-polarized gas at finite temperature \cite{rammelmuller_finite-temperature_2018} to solve for the chemical potential of both spins in the reservoir $\mu_{i\sigma}$. These determine the average chemical potential $\mu_i = (\mu_{i\downarrow} + \mu_{i\uparrow})/2$ and the Zeeman field $b_i = (\mu_{i\downarrow} - \mu_{i\uparrow})/2$ which, together with the temperature $T_i$, completely characterize the \newnew{universal pressure EoS $f_P(\beta\mu,\beta b)$} of the spatially homogeneous system
\begin{equation}
    P(\mu, b, T) = \frac{1}{\beta\lambda_T^3} f_P(\beta\mu, \beta b)
\end{equation}
where $\lambda_T = \sqrt{2\pi \hbar^2/m k_B T}$ is the thermal de Broglie wavelength and $\beta=1/k_B T$ is the inverse temperature. The free energy of a harmonically-trapped system $\Omega(\mu,b,T)$ can be computed from $P(\mu,b,T)$ (the free energy density) by applying the local density approximation with spatially-varying average chemical potential
\begin{equation}
    \mu(\vb{r}) = \mu - \frac{m}{2}(\omega_x^2 x^2 + \omega_y^2 y^2 + \omega_z^2 z^2)
\end{equation}
and spatially constant $b$ and $T$, yielding
\begin{align}
    -\Omega(\mu, b, T) &= \int\dd[3]{r} P[\mu(\vb{r}), b, T] \nonumber \\
    &= \frac{(k_B T)^4}{(\hbar\bar{\omega})^3} \frac{2}{\sqrt{\pi}} \int_0^\infty \dd{q} \sqrt{q} f_P(\beta\mu - q, \beta b) \nonumber \\
    &\equiv \frac{(k_B T)^4}{(\hbar\bar{\omega})^3} f_\Omega(\beta\mu, \beta b).
\end{align}
From \newnew{the universal EoS for the free energy $f_\Omega$} we can compute 
thermodynamic response functions like the compressibility $\kappa$, spin susceptibility $\chi$, and ``magnetic compressibility'' $\alpha$:
\begin{align}\begin{split}
    \kappa(\mu, b, T) &= \pdv{N}{\mu} = \frac{(k_B T)^2}{(\hbar\bar{\omega})^3} \pdv[2]{f_\Omega}{(\beta\mu)} \\
    \chi(\mu, b, T) &= \pdv{M}{b} = \frac{(k_B T)^2}{(\hbar\bar{\omega})^3} \pdv[2]{f_\Omega}{(\beta b)} \\
    \alpha(\mu, b, T) &= \pdv{N}{b} = \pdv{M}{\mu} = \frac{(k_B T)^2}{(\hbar\bar{\omega})^3} \pdv[2]{f_\Omega}{(\beta\mu)}{(\beta b)}
\end{split}\end{align}
which set the relationship between the atom number and magnetization imbalances and the chemical potential and Zeeman field biases in the linear response regime of the reservoirs
\begin{equation}
    \pmqty{\Delta N \\ \Delta M} \approx \frac{1}{2} \pmqty{\kappa & \alpha \\ \alpha & \chi} \pmqty{\Delta\mu \\ \Delta b}.
\end{equation}
A misalignment of the center of the magnetic trap with respect to the QPC leads to a finite atom number imbalance at equilibrium $\Delta N(\Delta\mu=0) = \Delta N_\infty \neq 0$. This can be compensated in the equation of state $\Delta\mu(\Delta N, N, T)$ by $\Delta\mu(\Delta N-\Delta N_\infty, N, T)$ where $\Delta N_\infty$ is fixed to $\Delta N$ at the longest transport time.

We find that all of our data lies within the computed range of the Zeeman field $\abs{\beta b} < 2$ (Sec.~\ref{sec:fitting_procedure}). Below the minimum computed chemical potential $\beta\mu<-5$, we use the third order virial expansion for $f_P(\beta\mu,\beta b)$ \cite{rammelmuller_finite-temperature_2018}. The densest regions of the cloud however lie above the critical degeneracy of $\beta\mu\approx2.5$ \cite{kuRevealingSuperfluidLambda2012} where the computed EoS is known to deviate from the true response of the system. We therefore take the EoS to be that of the low temperature superfluid with phononic excitations \cite{hou_first_2013}
\begin{equation}
    f_P^\mathrm{SF}(\beta\mu, \beta b) = \frac{16}{15\sqrt{\pi}}\bqty{\xi\pqty{\frac{\beta\mu}{\xi}}^{5/2} + \frac{\pi^4}{96}\pqty{\frac{3}{\beta\mu}}^{3/2}}
\end{equation}
where $\xi=0.370$ is the Bertsch parameter \cite{zurn_precise_2013}. The finite-temperature superfluid is known to have finite spin susceptibility \cite{long_spin_susceptibility_2021,rammelmuller_pairing_2021} so using this EoS, which has vanishing spin susceptibility $\pdv*[2]{f_P^\mathrm{SF}}{b} = 0$, introduces some systematic error. However, the susceptibility in this regime is exponentially suppressed by a large pairing gap so the overall error in $f_\Omega$ is negligible.

As the phase boundary between the normal and superfluid phases is not exactly known, we take it to be the intersection of $f_P$ and $f_P^\mathrm{SF}$ such that $f_P$ is continuous but its first derivative is discontinuous. This choice agrees well with a recent functional renormalization group approach to compute the phase boundary \cite{boettcher_phase_2015}. This choice also minimizes the potentially detrimental artifacts that could arise in the thermodynamic response functions which are derivatives of $f_\Omega$. In fact, there is no discontinuity in $\kappa$, $\chi$, or $\alpha$, likely because $f_\Omega$ is a weighted integral of $f_P$ and therefore naturally smoother.

\section{Data analysis}

\subsection{Estimate of the number of transport modes}
\label{sec:estimate_of_the_number_of_transport_modes}

An important parameter that contributes to the overall timescale of transport is the number of available transport modes in the QPC $\nch$ \cite{scheerSignatureChemicalValence1998}.  If the modes are de-coupled from each other and the current they each carry is identical, then the total current is $\nch$ times the current of a single mode $-(1/2)\dot{\Delta N}^K = \nch I_N^K$. In the non-interacting system, where it is possible to exactly compute the current in the Landauer formalism with our detailed knowledge of the potential energy landscape in the QPC \cite{corman_quantized_2019}, $\nch$ naturally emerges as the sum of the Fermi-Dirac occupation of each transverse mode indexed by the harmonic numbers $n_x, n_z$
\begin{equation}
    \nch = \sum_{n_x,n_z} \frac{1}{1 + \exp\qty{[E_{n_x,n_z}(0)-\mu]/k_B T}}
\end{equation}
where $E_{n_x,n_z}(0)$ is the energy of mode $n_x,n_z$ at the center of the QPC. This expression is valid if the mode transmits particles adiabatically (decoupled modes) \cite{ihnSemiconductorNanostructuresQuantum2010} which we have verified from our measurements on the non-interacting gas. The energy has contributions from the confinement along $x$ and $z$ from the two QPC beams and the gate beam as well as the residual repulsive light on the QPC that increases the energy. These two contributions are proportional to the square root of the power of the beam and linear in the beam power respectively. Finally including the gate potential $V_g$ from the attractive beam, the energy can be written as
\begin{align}\begin{split}
    E_{n_x,n_z}(0) =& \sqrt{(a_x\sqrt{P_x})^2 + \nu_{x,\mathrm{gate}}^2}(n_x+1/2) + b_x P_x \\ 
    + & a_z\sqrt{P_z}(n_z+1/2) + b_z P_z + V_g
    \label{eq:ZPE}
\end{split}\end{align}
where $P_{x/z}$ is the power of the QPC beam that confines along the $x/z$ direction and the parameters $a_x, b_x, a_z, b_z$ were calibrated by measuring $E_{0,0}(0)$ in a non-interacting system (the onset of quantized conductance) with various $P_x$ and $P_z$ and fitting Eq.~\ref{eq:ZPE}. Note that this estimation method assumes the same effective tunneling amplitude $\tau$ of each mode and neglects the possibility that the zero-point energy of the modes is reduced by the strong interparticle attraction (cf.~Ref.\cite{kanasz-nagy_anomalous_2016}).

\subsection{Estimate of the superfluid gap}
\label{sec:estimate_of_the_superfluid_gap}

In solid state systems, where the superconducting reservoirs and contacts are geometrically well-defined, the gap $\Delta$ that enters the theoretical model is given by the bulk of the reservoir material \cite{scheer_conduction_1997}. However, in our system where the separation between reservoirs, contact, and channel are not sharp but vary smoothly over the length scales of the Gaussian beams that define the QPC, this choice is not as obvious.

We therefore estimate $\Delta$ from the local chemical potential at the most degenerate location in the system at the contacts of the 1D region $\mu_c = \max_\mathbf{r}[\mu - V(\mathbf{r})]$ where $V(\mathbf{r})$ includes all traps, zero-point energies $E_{0,0}(\mathbf{r})$ of each confinement beam, and the gate potential $V_g(\mathbf{r})$. $\mu_c$ then gives the local density $n(\mu_c,T)=\pdv*{P}{\mu}$ from the equation of state of the homogeneous 3D system, and the Fermi energy $E_F = \hbar^2 (3\pi^2n)^{2/3}/(2m) \approx \mu_c/\xi$ determines the gap $\Delta = 0.44 E_F$ \cite{schirotzek_2008}. The temperature $k_BT/E_F < 0.04$ is low enough to safely apply the zero temperature value of $\Delta$. Note that this estimation method neglects the possibility that the additional confinement in the 2- and 1-dimensional regions enhances the pairing gap \cite{kanasz-nagy_anomalous_2016}.

\subsection{Fitting procedure}
\label{sec:fitting_procedure}
\begin{figure}
    \centering
    \includegraphics{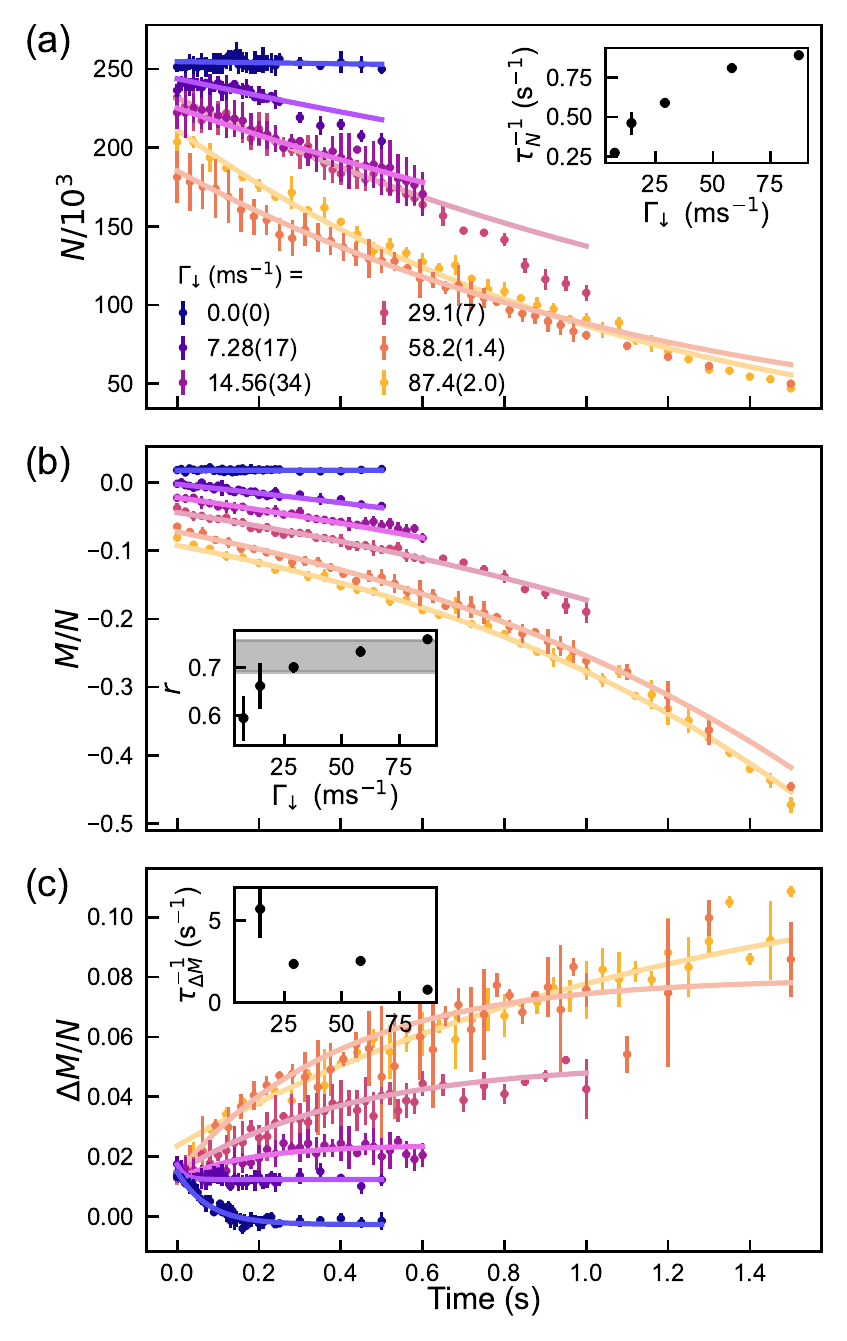}
    \caption{Atom number (a), magnetization (b) and relative magnetization (c) over time for varying dissipation strength. Data sets in (a) and (b) are vertically shifted for clarity by $+10$ and $-0.02$ units respectively where the $\Gamma_\downarrow=0$ data is the most shifted in (a) and unshifted in (b). The extracted fit parameters defined in Eq.~\ref{eq:loss_model} and \ref{eq:magnetization_model} are plotted in the insets. Inset of (a) is identical to the normalized atom loss rate plotted in Fig.~4(a). The horizontal bar in the inset of (b) represents the weighted average $r$ used in the fit of $\Delta N/N$. 
    }
    \label{fig:fig6}
\end{figure}

\new{\emph{Atom number decay.}---}
For a given data set (the time evolution of the density imbalance with a fixed configuration of the QPC, gate, and dissipation strength), we first fit the time evolution of the total atom number $N(t) = N_\downarrow(t) + N_\uparrow(t)$. As shown in the main text, the model cannot quantitatively predict the loss rates $\dot{N}_\sigma$ so we use a phenomenological model inspired by our previous work on a non-interacting gas under similar conditions \cite{corman_quantized_2019}. There, the loss rate $\dot{N}$ was found to be proportional to the fraction of atoms with energies above the zero point energy of the channel, which itself is approximately proportional to the atom number $N$. Moreover, we can assume that a $\kup$ atom is lost with some probability $r$ for each dissipated $\kdn$ atom, motivated by the dissipation mechanism described in Sec.~\ref{sec:dissipation_mechanism} where some fraction of the photon recoil is transferred from $\kdn$ to its paired $\kup$ via interactions. The parameter $r$ can therefore be interpreted as a measure of the spin correlations at the location of the dissipation. These two assumptions lead to the rate equations
\begin{align} \begin{split} \label{eq:loss_model}
    \dv{t} N(t) &= -\frac{1}{\tau_N} N(t) \\
    \dv{t} N_\uparrow(t) &= r \dv{t} N_\downarrow(t)
\end{split} \end{align}
with fit parameters $\tau_N$ and $r$ (the loss ratio) as well as the initial atom number in each spin state. We only fit the data before $\Delta N$ fully equilibrates ($\Delta N/N \geq 0.02$, on the order of our experimental stability of $\Delta N/N$) to obtain a better fit result at initial times, in which we are most interested. These fits for the data presented in Fig.~2 in the main text are shown in Fig.~\ref{fig:fig6}(a,b). We see that even this simple model fits the data well, allowing us to reliably extract $\dot{N}$ and $\dot{M}$ from the data.


\new{\emph{Full evolution including magnetization}---}
Now with the fitted $N(t)$ and $M(t)$, we turn to fitting $\Delta N(t)/N(t)$ and $\Delta M(t)/N(t)$ (we fit the relative imbalances $\Delta N/N$ and $\Delta M/N$ rather than the absolute imbalances $\Delta N$ and $\Delta M$ as this normalizes out the shot-to-shot atom number fluctuations of $\sim10\%$). Using the EoS for the half-harmonic reservoirs described in Sec.~\ref{sec:reservoir_thermodynamics}, the measured atom number $N_i(t)$ ($i=L,R$), magnetization $M_i(t)$, and temperature $T_i(t)$ in each reservoir at a given time $t$ can be converted to a chemical potential $\mu_i(t)$ and Zeeman field $b_i(t)$. The biases between the reservoirs $\Delta\mu=\mu_L-\mu_R$ and $\Delta b=b_L-b_R$ and the average temperature $T=(T_L+T_R)/2$, along with the superfluid gap $\Delta$ and number of modes $\nch$ (both determined by the average chemical potential $\mu = (\mu_L + \mu_R)/2$ and the gate potential correction factor $\eta_g$), tunneling amplitude $\tau$, and dissipation rates $\gamma_\downarrow$ and $\gamma_\uparrow$ are then the inputs to our model to compute the density and spin currents $I_N^K$ and $I_M^K$ (Sec.~\ref{sec:theory}). In this way, the decay of the density and magnetization imbalances is described by the system of non-linear differential equations
\begin{align}\begin{split}
    -\frac{1}{2}\dv{t} \Delta N^K &= \nch I_N^K(\Delta N^K, \Delta M^K, N, M, T, \eta_g, \tau, \gamma_\downarrow, \gamma_\uparrow) \\
    -\frac{1}{2}\dv{t} \Delta M^K &= \nch I_M^K(\Delta N^K, \Delta M^K, N, M, T, \eta_g, \tau, \gamma_\downarrow, \gamma_\uparrow)
\end{split}\end{align}
subject to the initial conditions $\Delta N^K(0) = \Delta N(0)$ and $\Delta M^K(0) = \Delta M(0)$, which can be easily solved numerically. However, we find that for values of $\gamma$ that reproduce the observed $I_N$, the computed $I_M^K$ underestimates the observed $I_M = -\dot{\Delta M}/2$ by over an order of magnitude. This means that we cannot obtain good fits by simultaneously fitting $\Delta N(t)/N(t)$ and $\Delta M(t)/N(t)$.

The evolution of the relative magnetization $\Delta M/N$ is shown in Fig.~\ref{fig:fig6}(c) along with a fit to a phenomenological rate equation assuming linear spin transport
\begin{equation} \label{eq:magnetization_model}
    \dv{t} \pqty{\frac{\Delta M(t)}{N(t)}} = -\frac{1}{\tau_{\Delta M}}\bqty{\frac{\Delta M(t)}{N(t)} - \pqty{\frac{\Delta M}{N}}_\infty}.
\end{equation}
Similarly to how a chemical potential bias $\Delta\mu$ can drive a magnetization current $I_M$ due to the spin asymmetry $\gamma_\downarrow - \gamma_\uparrow$, the Zeeman field bias induced by the magnetization imbalance $\Delta b \approx \Delta M/\chi$ can drive a particle current $I_N$. Furthermore, $\Delta M$ can directly influence $\Delta\mu$ via the ``magnetic compressibility'' $\Delta\mu \approx \Delta N/\kappa - (\alpha/\kappa\chi)\Delta M$ (Sec.~\ref{sec:reservoir_thermodynamics}). Because the model cannot accurately reproduce the off-diagonal conductance $\pdv*{I_M}{\Delta\mu}$, we cannot expect it to compute $\pdv*{I_N}{\Delta b}$ either. However, on general grounds, we can expect that $\pdv*{I_N}{\Delta b} \approx \pdv*{I_M}{\Delta\mu}$---essentially a statement of Onsager's reciprocal relations for coupled density and spin transport. Our data shows that, for the strongest dissipation, $\pdv*{I_M}{\Delta\mu} \sim -\nch/h$ and therefore the contribution of $\Delta b$ to $I_N$ is $\lesssim-\nch\Delta b/h$. The $\Delta b$ extracted from $\Delta M$ shown in Fig.~\ref{fig:fig6}(c) is $\Delta b/\Delta < 0.03$ and is far below this limit for all but the longest transport times at the strongest dissipation, so the influence on $I_N$ is below even the normal part of the chemical-potential driven transport $\sim\nch\Delta\mu/h$, let alone the superfluid part $\sim\nch\Delta/h$. Furthermore, the influence on $\Delta\mu$ driven by $\Delta M$ is $(\alpha/\kappa\chi) \Delta M < 0.005\Delta$ in the most extreme case which is negligible compared to the density imbalance-induced bias $\Delta N/\kappa \sim 0.05\Delta$. In summary, the influence of the magnetization dynamics $\Delta M(t)$ on the density dynamics $\Delta N(t)$ is negligible to a good approximation.

\new{\emph{Simplified evolution of density imbalance.}---}Therefore, because $\Delta N(t)$ is decoupled from $\Delta M(t)$ and $I_N^K$ is insensitive to $\gamma_\downarrow-\gamma_\uparrow$, we fit only $\Delta N(t)/N(t)$ with our model for $\Delta N^K(t)$ using the spin-balanced equation of state ($b=\Delta b=0$) and with the ratio of the dissipation strengths fixed to the fitted loss ratio $\gamma_\uparrow/\gamma_\downarrow = r = \dot{N}_\uparrow/\dot{N}_\downarrow$
\begin{equation}
    -\frac{1}{2}\dv{t} \Delta N^K = \nch I_N^K(\Delta N^K, N, T, r, \eta_g, \tau, \gamma_\downarrow).
\end{equation}
As explained in the main text, we first fit the imbalance decay curves where the dissipation beam was not present such that we can fix $\gamma_\downarrow=\gamma_\uparrow=0$. The fit parameters here are the average tunneling amplitude of the occupied modes $\tau$ and, due to its experimental uncertainty (Sec.~\ref{sec:defining_the_quantum_point_contact}), the correction factor $\eta_g$ of the gate potential $V_g$ which enters into $\nch$ and $\Delta$. The fitted values of $V_g$ are consistent with its calibrated value for the data shown and within 20\% for other measurements with different QPC parameters ($\nu_x$ and $V_g$), supporting this approach. Although the fitted $r$ shows a small dependence on $\Gamma_\downarrow$, we simply use the average value for the fit of $\Delta N/N$ due to the uncertainty in the fit of $r$ and the fact that the calculated current is insensitive to $r$. With $r$, $\tau$ and $\eta_g$ fixed, we then fit $\gamma_\downarrow$ (which determines $\gamma_\uparrow=r\gamma_\downarrow$) to each data set taken under the same QPC and gate conditions but with varying dissipation strength.


\subsection{Numerical derivation of current}

To compare the data with the theory in current-bias relation  [Fig.~2(b)], we obtain the current $I_N$ from a numerical derivative of the measured $\Delta N$. In practice, we pass the relative imbalance data $\Delta N(t)/N(t)$ [Fig.~2(a)] into a Savitzky–Golay filter of order 3, from which the first derivative at each time point is given by the 4 neighbouring points. $\dot{\Delta N}$ is then obtained assuming the fitted exponential decay of $N$. To estimate the uncertainty of the numerical time derivative, we apply a bootstrap-type method. We pass an averaged time evolution with a random sampling without replacement of 70\% of the entire dataset to the Savitzky-Golay filter to have an ensemble of derivative results (100 realizations). We then take the standard deviation of this ensemble to estimate its uncertainty. There is however still a visible oscillation in the derivative data beyond the estimated uncertainty, which is a typical artifact due to the limited sampling in time in the original data.   

\subsection{Removing outliers}

Due to some technical issues such as atom number fluctuation and sporadic synchronization failure of our spatial light modulator that generates the dissipative beam, we have data points that are clearly beyond statistical uncertainties. We remove the outliers beyond 3.5$\sigma$ of the statistical distribution of $\Delta N/N$ of the entire dataset in a given channel and dissipation configuration, and those beyond 3$\sigma$ in $N$, $N_\downarrow/N_\uparrow$, and $\Delta M/N$ which are better indicators of technical problems. As we have different transport time in the dataset, we use a smoothing spline to determine the mean value of the data as function of the transport time. This procedure assumes no knowledge of our theoretical model.

\end{document}